\documentstyle[12pt]{article}

\begin{document}

\def\SG{\mbox{\Large$\sigma$\normalsize}}

\def\a#1#2#3#4{{\,{#1}_{#3}{#2}_{#3}\cdots\,{#1}_{#4}{#2}_{#4}}}

\def\an#1#2{\a{#1}{#2}{1}n}

\def\vc#1#2#3{{\,{#1}_{#2}\cdots\,{#1}_{#3}}}

\def\Va#1#2#3#4#5#6{\underline{\a{#1}{#2}{#3}{#4}}\a{#1}{#2}{#5}{#6}}

\def\va#1#2#3{\Va{#2}{#3}{1}{#1}{#1+1}{n}}

\def\vk#1#2#3#4#5{\underline{\vc{#1}{#2}{#3}}\vc{#1}{#4}{#5}}

\def\ind#1#2#3{^{\phantom{#1} #2}_{#1 \phantom{#2} #3}}

\def\ra#1{{#1\llap{/}}}

\def\pts#1#2{\; {}^{{}^{#1}}\llap{.\,.}\rlap{\,.}{}^{{}^{#2}} \;}

\def\raya#1#2#3#4{{\,{#1}_{#2}\cdots\,{\ra#1}_{#3}\cdots\,{#1}_{#4}}}

\def\a#1#2#3#4{{\,{#1}_{#3}{#2}_{#3}\cdots\,{#1}_{#4}{#2}_{#4}}}

\def\an#1#2{\a{#1}{#2}{1}n}

\def\vc#1#2#3{{\,{#1}_{#2}\cdots\,{#1}_{#3}}}

\def\Va#1#2#3#4#5#6{\underline{\a{#1}{#2}{#3}{#4}}\a{#1}{#2}{#5}{#6}}

\def\va#1#2#3{\Va{#2}{#3}{1}{#1}{#1+1}{n}}

\def\vk#1#2#3#4#5{\underline{\vc{#1}{#2}{#3}}\vc{#1}{#4}{#5}}

\def\c#1#2{\mbox{\scriptsize$\left(\begin{array}{c} #1 \\ #2 \end
  {array}\right)$\normalsize}}

\def\ra#1{{#1\llap{/}}}

\def\pts#1#2{\; {}^{{}^{#1}}\llap{.\,.}\rlap{\,.}{}^{{}^{#2}} \;}

\def\ind#1#2#3{^{\phantom{#1} #2}_{#1 \phantom{#2} #3}}

\def\raya#1#2#3#4{{\,{#1}_{#2}\cdots\,{\ra#1}_{#3}\cdots\,{#1}_{#4}}}

\def\OMEGA{\mbox{\large$\Omega$\normalsize}}

\def\SIGMA{\mbox{\Large$\sigma$\normalsize}}

\def\base{{\cal S}}

\def\LL{{\cal L}}
\title{The Extended Loop Group: \\
an infinite dimensional manifold associated \\
with the loop space }

\author{ \\ \\ \small
{\bf Cayetano Di Bartolo} \\
\small Departamento de F\'\i sica, Universidad Sim\'on Bol\'\i var, \\
\small Apartado 89000, Caracas 1080-A, Venezuela.
\and    \\
\small {\bf Rodolfo Gambini and Jorge Griego }\\
\small Instituto de F\'{\i}sica, Facultad de Ciencias,\\
\small Trist\'an Narvaja 1674, Montevideo, Uruguay.}

\date{February, 1993}
\maketitle
\vspace{0.5cm}

\begin{abstract}
     A set of coordinates in  the  non  parametric  loop-space  is
introduced. We show that these coordinates transform under infinite
dimensional   linear    representations of   the    diffeomorphism
group. An extension of the  group  of  loops  in  terms  of  these
objects is proposed. The
enlarged group behaves locally  as  an  infinite  dimensional  Lie
group. Ordinary loops form a subgroup of this  group. The  algebraic
properties of this new  mathematical  structure  are  analized  in
detail. Applications  of  the  formalism  to  field   theory, quantum
gravity and knot theory are considered.
\end{abstract}


\newpage

\section{\bf Introduction}

Loop space has been used in several non-perturbative approaches to gauge
theories and gravitation\cite{R1,R2,R3,R4,R5,R6,R7}. In the eighties, the
loop representation of gauge theories was accomplished\cite{R8,R9,R10,R11}.
This representation has proved to be a suitable framework where one can
develop a complete canonical scheme for the quantization of gauge theories.
The two remarkable features of this formulation are the manifest gauge
invariance of the quantization method and the solution of the constraints
through geometrical requirements. Other approaches to the loop space have
also been developed, based fundamentally on the Polyakov and
Makeenko-Migdal 's\cite{R4,R5,R12} treatment of the dynamics of loop
dependent objects in Yang-Mills theories. At present, the results obtained
in the loop representation are in agreement with those obtained in the
usual approach to gauge theories. The loop representation also gives new
insights into the non perturbative aspects of gauge theories.

Recently, the loop representation has been used in quantum general
relativity\cite{R13,R14}. This representation emerges naturally from the
Ashtekar's new formulation of general relativity\cite{R15}. Any quantum
field theory whose configuration variable is a connection can be
realized in the loop space language. But in the quantum gravity case, the
use of the loop space turns out to be an essential tool in order to develop
a complete non perturbative quantization program \cite{R16}. Non
perturbative solutions to all the constraints of quantum gravity have been
found\cite{R17,R18}, including those associated with nondegenerate
metrics\cite{R19}. In the latter case, knot invariants of
intersecting loops have proved to play an essential role in the
construction of the solutions.

Knot and quantum field theories can be related in different ways. For
instance, by making use of the deep connection between the three
dimensional Chern-Simons models and certain two dimensional conformal field
theories, Witten\cite{R20} was able to derive a skein relation for the
Wilson line expectation value, $<W(L)>$, and show that there exists a
correspondence between $<W(L)>$ and Jones\cite{R21} and HOMFLY\cite{R22}
polynomials. Also, the construction of link polynomials from exactly
solvable models in statistical mechanics was developed\cite{R23}. On the
other hand, as we just mentioned, knot invariants are intimately related
with the non perturbative quantum states of general relativity. Actually,
at this stage one can speak about a loop space formalism. This formalism
seems to be an appropriate framework to develop a realistic quantum theory
of space time\cite{R16}.

In the loop representation, states are given as functionals of loops. Loops
can be precisely defined in a non parametric form\cite{R9} as an
equivalence class of closed oriented paths in a manifold. We will declare
the loop L equivalent to L' if the product $L' \circ \overline L$, the
contour obtained by following first L' and then L with the opposite
orientation, is a closed path which is contractible within itself to the
null path. The basic mathematical property that makes loop space an
appropriate scenario to construct representations of gauge theories and
gravitation is the group property. Loops form a group and the holonomy of a
particular gauge theory may be obtained by considering a representation of
the group of loops into the particular group being gauged.

The group of loops does not contain one parameter subgroups and
consequently, it is not a Lie group. However, by analyzing its infinitesimal
elements one can appropriately introduce generators (through the loop
derivative operator) and construct with them any element of the group. The
group composition law is the fundamental mathematical property that
enables us to operate in loop space and to use it as the domain of
functionals where the gauge theory is realized.

The lack of a Lie structure for the loop group restricts our capability to
work with the loop space. We are not dealing with a manifold and then we
can not use all the functional techniques available
in those cases. From a mathematical point of view, it would be highly
desirable to develop an extension of the group of loops in order to include
it in a larger manifold structure.

 From a physical point of view, the fact that loops are non local
distributional objects with a one dimensional support introduces certain
typical difficulties in the hamiltonian formulation of any diffeomorphism
invariant gauge theory in the loop representation. It is a well known fact
that many of the diffeomorphism invariant quantities up to now considered
(i. e. the Gauss number) are ill defined when evaluated on loops and need
to be framed. Moreover, the framing destroys invariance under
diffeomorphism of the theory. Another typical difficulty is related with
our inability to define an integration measure in loop space
and consequently, an inner product in the Hilbert space of quantum gravity.
We expect that the extension of the loop space would allow to transform
loop integrals into functional integrals.

In this paper we develop a new approach to the loop space formalism. The
basic idea underlying this approach is the possibility to define a {\it
coordinate representation} on the loop space. The coordinates on loop space
were first introduced by Gambini and Leal\cite{R24} and were used by
Br\"ugmann, Gambini and Pullin in the determination of
the first nondegenerate quantum state of general relativity\cite{R19}.

We start by introducing a set of coordinates which allows to describe any
element of the non parametric loop space. We show that this coordinates
transforms under linear representations of the diffeomorphism group.
We find that it is possible to generate an infinite set of gauge invariant
objects generalizing the coordinate expansion of the holonomy. The
identification of a set of invariant tensors under diffeomorphism
transformations allows to give rules for the construction of an, in
principle infinite, set of knot invariants. We also show how to generate
new knot invariants from others and how to relate link and knot invariants
in the SU(2) case. Within this approach, the basic conditions necessary to
define an affine geometry in loop space seem to be  fulfilled.What
is more important,loop coordinates allow to show that
there exist an infinite dimensional manifold with a local Lie group
structure associated with the loop space, the Special-extended Loop Group.
This group provides the basis of a new formulation of the loop
representation. A complete description of the algebraic properties of this
enlarged structure is accomplished. Finally we would like to stress
that   we  are  not  going  to  discuss  analytical   details   of
regularization and, in general, this
presentation  pretend  a physicist's rather  than  a
mathematician's level of rigor.

We organize this article as follows: in section 2 we introduce a set of
coordinates on loop space and show that they are connected by infinite
dimensional linear transformations under diffeomorphisms. We also show that
these coordinates are not independent, but satisfy a set of algebraic and
differential constraints. Section 3 is devoted to the differential
constraint. In section 3.1 we solve the differential constraint and in
section 3.2 non trivial representations of the diffeomorphism group (with a
fixed point) are introduced. The
free coordinates associated with loops are
defined in section 4. In section 5 invariant tensors in the space of free
coordinates are introduced and the relationship between the invariant
tensors and knot invariants  is  analyzed.The  first  five sections
lays the groundwork for the further study of the extended loop group.
In section 6 we show how the loop coordinates can
be incorporated in a more general algebraic structure, the SeL group. The
algebra of the SeL group is studied in section 7. The generators are
introduced in Section 7.1, in section 7.2 a basis for the algebra is
constructed and section 7.3 is devoted to the study of invariants forms and
its relationship with  the  automorphism  transformations  of  the
algebra. Final comments and conclusions are made in section 8.

\section{\bf Coordinates on the loop space}

The loop representation\cite{R8,R13} of a quantum field theory can be
constructed
in terms of an algebra of linear operators defined on a state space of
loop functionals. These states may be expressed as follows:
\begin{equation}
\Psi (L_1,\ldots,L_n) = \int d\mu [A] \, \Psi [A] \, W_A(L_1)\cdots W_A(L_n)
\label{psi}
\end{equation}

\noindent where
\begin{equation}
W_A(L) = Tr\left[ P \, \exp \oint_L A_a\, dy^a \right] = Tr[ U_A(L) ]
\label{wilson}
\end{equation}

\noindent
is the Wilson line functional corresponding to the holonomy $U_A(L)$. Thus,
holonomies are the building blocks of any loop dependent object. The
holonomy of a particular gauge theory\cite{R9} may be obtained by
considering a representation of the group of loops into the particular
group G being gauged. For Quantum Gravity $U_A(L)$ belongs to
SU(2)\cite{SU2}. As it has been emphasized by a number of
authors\cite{R9,R13,R25,R26,R27,R28}, loops are equivalence classes of
closed oriented paths in a manifold M.
As it was already emphasized in the introduction,the set of loops forms a
group. It is
important to note that whereas closed curves are essentially
parametrization dependent objects, loops may be described in nonparametric
form\cite{R10,R29}. This means that the space of loops on which the loop
representation is defined is not the standard parametrized space of closed
curves.

In this section we are going to introduce a set of coordinates which allows
to describe any element of the nonparametric loop space. Then, we are going
to show that the corresponding coordinates of two diffeomorphic loops are
connected by an infinite dimensional linear transformation. In other words,
the loop coordinates transform under linear representations of the
diffeomorphism group.

Let us start by introducing a set of coordinates in loop space. They are
related with the holonomies as follows

\begin{equation}
U_A(L)=1+\sum_{n=1}^{\infty} \int dx^3_1\cdots dx^3_n
       A_{a_1}(x_1)\cdots A_{a_n}(x_n) X^{\vc a1n}(x_1,\ldots ,x_n,L)
\label{holonomia}
\end{equation}

\noindent
where L are loops with a base point O taken as their origin and the loop
dependent objects X are given by:

\begin{eqnarray}
X^{\vc a1n}(x_1,\ldots,x_n,L) &=& \int_Ldy_n^{a_n}
   \int_0^{y_n}dy_{n-1}^{a_{n-1}}\cdots \int_0^{y_2}dy_{1}^{a_1}
   \delta (x_n-y_n)\cdots \delta(x_1-y_1) \nonumber
\\ & \label{X} \\
&=&\oint_Ldy_n^{a_n}\cdots \oint_Ldy_{1}^{a_1}
   \delta (x_n-y_n)\cdots \delta(x_1-y_1) \Theta_L(0,y_1,\ldots,y_n)
\nonumber
\end{eqnarray}

\noindent
and $\Theta_L(0,y_1,\ldots,y_n)$ orders the points along the contour
starting at the origin of the loop. These relations define the X objects of
"rank" n. We shall call them the multitangents of the loop L.

In what follows, it will be convenient to introduce the notation

\begin{equation}
X^{\vc \mu 1n}(L) = X^{\a ax1n}(L) = X^{\vc a1n}(x_1,\ldots,x_n,L) \;,
\end{equation}

\noindent
with $\mu_i \equiv (a_ix_i)$, which is more suggestive of the role played
by the x variables under diffeomorphic transformations.

The distributional objects X allow to determine the Wilson line element
(\ref{wilson}) for any connection. As it was first pointed out by Makeenko
and Migdal\cite{R30}, the wave functions in the loop representation depend
on the cyclic permutations of the multitangent fields, given by

\begin{equation}
X_c^{\vc \mu 1n}(L) = \frac {1}{n} \; \bigl\{ \, X^{\vc \mu 1n}(L)
         +X^{\mu_2\cdots \mu_n \mu_1}(L) + c.p. \, \bigr\}
\end{equation}

The X objects transform as multivector densities under the subgroup of
coordinate transformations that leaves the base point O fixed. In other
words if

\begin{equation}
x^a \longrightarrow x'^a=D^a(x)
\end{equation}

\noindent
then

\begin{equation}
X^{\an a{x'}}(DL) = \frac{\partial {x'}_1^{a_1}}{\partial x_1^{b_1}}
 \cdots \frac{\partial {x'}_n^{a_n}}{\partial x_n^{b_n}}
   \frac 1{J(x_1)}\cdots \frac 1{J(x_n)} X^{\an bx}(L) \,.
\label{ley}
\end{equation}

\noindent
where $J$ is the jacobian of the transformation.

Introducing a matrix notation this relation may be rewritten in a different
way which is more suggestive of the role of the multitangents in loop
space. Let us define the vector-like object $\vec{X}(L) \equiv
(X^{\mu_1}(L), \cdots ,X^{\vc \mu 1i}(L), \cdots \, )$. The components of
this object are multivector density fields of any rank (being the rank
the number of indexes). Now we introduce the matrix $\Lambda_D$ with
components

\begin{equation}
\Lambda\ind{D}{\vc \mu 1n}{\vc \nu 1m} \; \equiv \;
\delta_{n,m} \;\,
\Lambda\ind{D}{\mu_1}{\nu_1} \cdots \, \Lambda\ind{D}{\mu_n}{\nu_n}
\label{lD}
\end{equation}

\noindent
where

\begin{equation}
\Lambda_{D \phantom{ay} bx}^{\phantom{D} ay} = \frac 1{J(x)}
\frac{\partial D^a(x)}{\partial x^b} \;\delta \bigl(x-D^{-1}(y)\bigr)
 = \frac{\partial D^a(x)}{\partial x^b} \;\delta \bigl(D(x)-y)\bigr) \;.
\end{equation}

\noindent
Then (\ref{ley}) can be written in the form

\begin{equation}
X^{\vc \mu 1n}(DL) = \sum_{m=1}^\infty
\Lambda\ind{D}{\vc \mu 1n}{\vc \nu 1m} \, X^{\vc \nu 1m}(L)
\end{equation}

\noindent
or shorthanded

\begin{equation}
\vec{X}(DL) \; = \; \Lambda_D \; \cdot \; \vec{X}(L) \;.
\end{equation}

\noindent
where use has been made of a generalized Einstein convention for the
repeated indexes, given by

\begin{equation}
A_{bx}\;B^{bx} \equiv \sum^3_{b=1} \int\,d^3x A_{bx}\;B^{bx} \, .
\end{equation}

\noindent
Hence, the multitangents associated with two diffeomorphic loops are
related by linear transformations and therefore they transform as
generalized tensors. The constant (loop independent) linear transformations
are the elements of a linear representation of the diffeomorphism group. In
certain sense it is a trivial linear representation of the diffeomorphism
group because it is a direct consequence of the multivector character of
the multitangent field. Later on, we shall introduce a new non-trivial
representation which is relevant in the determination of the knot invariant
quantities.

The multitangents contain all the relevant information necessary to
uniquely determine a loop and therefore they may be considered as good
prospects for coordinates of a loop. However, they are not independent
variables. In fact, they obey a set of simple algebraic and differential
constraints.

The algebraic constraints are a direct consequence of the properties of the
$\Theta_L$ functions.

\begin{eqnarray}
&&\Theta_L(0,y_1) = 1 \;,\phantom{AAAAAAAAAA}
 \Theta_L(0,y_1,y_2) + \Theta_L(0,y_2,y_1) = 1 \;, \nonumber
\\ && \\
&&\Theta_L(0,y_1,y_2,y_3) +\Theta_L(0,y_2,y_1,y_3) + \Theta_L(0,y_2,y_3,y_1)
 = \Theta_L(0,y_2,y_3) \nonumber
\end{eqnarray}

\noindent
and so on. The corresponding constraints for the X objects are

\begin{eqnarray}
&&X^{\mu_1} = X^{\mu_1}, \phantom{AAAAA}
X^{\mu_1 \mu_2} + X^{\mu_2 \mu_1} = X^{\mu_1} \,X^{\mu_2},
\nonumber \\ && \\
&&X^{\mu_1 \mu_2 \mu_3} + X^{\mu_2 \mu_1 \mu_3} +
X^{\mu_2 \mu_3 \mu_1} = X^{\mu_1} \,X^{\mu_2 \mu_3} \, .
\nonumber
\end{eqnarray}

\noindent
Their general form is

\begin{equation}
X^{\vk \mu 1k{k+1}n} \equiv
\sum_{P_k} X^{P_k(\vc \mu 1n)} = X^{\vc\mu 1k} \,X^{\vc\mu{k+1}n}
\label{VA}
\end{equation}

\noindent
and the sum goes over all the permutations of the $\mu$ variables which
preserve the ordering of the ${ \mu_1, \ldots, \mu_k }$ and the ${
\mu_{k+1}, \ldots, \mu_n }$ between themselves.

The differential constraint may be easily obtained from Eq.(\ref{X}) and
takes the form:

\begin{equation}
\frac {\partial \phantom{iii}} {\partial x_i^{a_i}} \,
X^{\a ax1i\,\ldots\,a_nx_n} =
 \bigl( \,\delta(x_i-x_{i-1}) - \delta(x_i-x_{i+1}) \,\bigr)
 X^{\a ax1{i-1}\,\a ax{i+1}n}
\label{dc}
\end{equation}

\noindent
with $x_0$ and $x_{n+1}$ equal to the origin of the loop.

The differential constraint is related with the gauge transformation
properties of the holonomy. Any object $\vec{E} = (E^{\mu_1}, \cdots
,E^{\vc \mu 1n}, \cdots \, )$ satisfying the differential constraints
allows one to define a gauge covariant quantity

\begin{equation}
U_A(E) = 1+\sum_{n=1}^{\infty} \int dx_1\cdots dx_n
       A_{\mu_1}\cdots A_{\mu_n} E^{\vc\mu 1n} \,.
\label{holoE}
\end{equation}

Even though for any loop there corresponds a distributional object $
\vec{X}(L)$ such that (\ref{holoE}) generates their Wilson functional, the
converse is not true. The space of functions that satisfy both
algebraic and differential constraints is more general and includes the
multitangents $\vec{X}(L)$ and smooth functions. For simplicity, one can
consider the abelian case. For U(1), loops are completely described through
the multitangent fields of rank one. We know that $X^{\mu_1}(L)$ is given
by integration of a distribution along the loop L. This function is
transverse (divergence free) and trivially satisfy the algebraic
constraint. But this set of functions is only a sector of the entire space
of transverse functions.

\section{\bf The differential constraint}

\subsection{\bf The solution of the differential constraint}

We consider now the solution of the differential constraint. Let us start
by introducing transverse and longitudinal projectors in the multivector
density space. We shall first consider a covariant metric in the space of
transverse vector densities of rank one. Given two transverse fields
$V^{ax}$ and $W^{ax}$ one can define\cite{R31}

\begin{eqnarray}
&&g(V,W) = \int d^3 x \phantom{ii} V^aA_a^W \nonumber
\\ && \label{metrica} \\
&&\partial _aV^a = \partial _aW^a = 0 \nonumber
\end{eqnarray}

\noindent
where $A_a^W$ is the potential associated with the curl free tensor $W_{ab}
= \epsilon _{abc} W^c = \partial _{[a} A_{b]}^W$.

In the transverse (non covariant) gauge

\begin{equation}
\partial ^a A_a^W = 0
\end{equation}

\noindent
it takes the form

\begin{equation}
g(V,W) = g_o{}_{\; axby} \, V^{ax} W^{by}
\end{equation}

\noindent
with

\begin{equation}
g_o{}_{\; axby} \; = \;- \epsilon_{abc} \,\, \frac {\partial^c} {\Delta} \,
\delta(x-y)     \; = \;- {\frac {1}{4\pi}} \; \epsilon_{abc} \; \frac
{x^c-y^c} {\mid x-y \mid ^3} \;\;.
\end{equation}

\noindent
This well known object is the kernel of the Gauss knot invariant. It is the
expression in a particular gauge of the covariant metric in the space of
transverse vector densities defined by (\ref{metrica}). In general, the
covariant metric is defined up to gradients

\begin{equation}
g_{axby} = g_o{}_{\; axby} + \rho_{ax \, y,b} + \rho_{by \, x,a} \;.
\label {g}
\end{equation}

\noindent
Now transverse and longitudinal projectors may be easily written in terms
of g and its inverse in the transverse space

\begin{equation}
g^{axby} = \epsilon^{abc} \,\, \partial_c \,\delta(x-y) \;.
\label{ginverse}
\end{equation}

\noindent
In fact

\begin{equation}
\delta\ind{T}{ax}{by} \equiv g^{ax\,cz} g_{cz\,by}
\end{equation}

\noindent
and

\begin{equation}
\delta\ind{L}{ax}{by} \equiv \delta\ind{}{ax}{by}- \delta\ind{T}{ax}{by}
\equiv \delta_{a,b} \, \delta (x-y) - \delta\ind{T}{ax}{by}
\label{deltaL}
\end{equation}

\noindent
are orthogonal projectors

\begin{eqnarray}
&&\delta\ind{T}\mu\rho \;\; \delta\ind{T}\rho\nu \, = \,
\delta\ind{T}\mu\nu \;, \nonumber
\\&& \nonumber \\
&&\delta\ind{L}\mu\rho \;\; \delta\ind{L}\rho\nu \, = \,
\delta\ind{L}\mu\nu \;, \nonumber
\\ && \nonumber \\
&&\delta\ind{L}\mu\rho \;\; \delta\ind{T}\rho\nu \, = \,
\delta\ind{T}\mu\rho \;\; \delta\ind{L}\rho\nu \, = \, 0 \;.
\nonumber
\end{eqnarray}

By using the explicit form of the covariant metric, one can prove that

\begin{equation}
\delta_{L \phantom{AA} by}^{\phantom{A} ax} \, = \,
\phi_{ \phantom{AA} y,b}^{ax}
\end{equation}

\noindent
where

\begin{equation}
\frac {\partial} {\partial x^a} \; \phi_{\phantom{AA} y}^{\,\, ax} \, = \,-
\delta(x-y) \;.
\label{cond}
\end{equation}

The ambiguity in the definition of the metric induces an ambiguity in the
decomposition in transverse and longitudinal parts. Each function $\phi$
satisfying (\ref{cond}) determines a particular prescription of the
decomposition. It is important to note that the transverse density fields
and in particular the contravariant metric (\ref{ginverse}) are
prescription independent. When the transverse metric $g_o$ is chosen we
have

\begin{eqnarray}
\phi_{o \phantom{Ai} y}^{\; ax} \; &=& \; {\frac {1}{4\pi}} \; \frac
{\partial}
{\partial x^a} \frac {1}{\mid x-y \mid}
\\ && \nonumber \\
\delta_{oT \phantom{A} by}^{\phantom{Ai} ax} \; &=& \;
\delta_{\phantom{AA} by}^{ax} + \frac {\partial^a \partial_b} {4\pi} \frac
{1}{\mid x-y \mid} \;\;.
\end{eqnarray}

Now we are ready to solve the differential constraint. Consider the set
${\cal E}^*$ of all quantities $\vec{E}$ that satisfy the differential
constraint and whose components $E^{\vc\mu 1n}$ are multivector density
fields. This set forms a linear vector space. A transverse projector acting
on the vector space ${\cal E}^*$ can be introduced through the matrix
$\delta_T$, defined as

\begin{equation}
\delta\ind{T}{\vc\mu 1n}{\vc\nu 1m} \; \equiv \;
\delta_{n,m} \;\;
\delta\ind{T}{\mu_1}{\nu_1}\cdots \delta\ind{T}{\mu_n}{\nu_n} \;.
\end{equation}
The transverse part of any element of ${\cal E}^*$ is given by

\begin{equation}
\vec{Z} \; = \; \delta_T \; \cdot \; \vec{E} \;.
\label{Z}
\end{equation}

The vector $\vec{Z}$ satisfies the homogeneous differential constraints.
The set of all $\vec{Z}$'s forms a linear vector space ${\cal Z}$. Equation
(\ref{Z}) may be inverted allowing to write $\vec{E}$ in terms of
$\vec{Z}$. To
do that, we start evaluating

\begin{equation}
E^{\vc\mu 1n} =
\delta\ind{}{\mu_1}{\nu_1}\cdots \delta\ind{}{\mu_n}{\nu_n}
E^{\vc\nu 1n}
\end{equation}

\noindent
and making use of the decomposition of the identity (\ref{deltaL}), the
differential constraint and recalling that $E^{\mu_1} = Z^{\mu_1}$, we get

\begin{equation}
\vec{E} \;= \; \SG \; \cdot \; \vec{Z}
\label{ESZ}
\end{equation}

\noindent
where the {\it soldering quantities} $\SG$ only depend on the function
$\phi$ which characterize the choice of decomposition in transverse and
longitudinal parts,

\begin{equation}
\SG\ind{}{\vc\mu 1n}{\vc\nu 1m} = \left\{
\begin{array}{ll}
\delta\ind{T}{\vc\mu 1n}{\vc\nu 1n}\;, & \mbox{ if $m = n$} \rule{0mm}{6mm}
\\ \vphantom{ Q^{{}^1}_{{}_1} }
Q^{\vc\mu 1n}_{\vc\rho 1{n-1}}
\,\SG\ind{}{\vc\rho 1{n-1}}{\vc\nu 1m}\;,
&\mbox{ if $m < n$} \rule{0mm}{6mm} \\ \rule{0mm}{6mm}
\\
0\;, &\mbox{ if $m > n$}
\end{array}
\right.
\end{equation}

\noindent
with

\begin{equation}
Q^{\an ax}_{\a cy1{n-1}} \equiv \sum^n_{j=1}
\delta^{\a ax1{j-1}}_{\a cy1{j-1}}
\left(\phi^{a_jx_j}_{\,y_j}-\phi^{a_jx_j}_{\,y_{j-1}}\right)
{\delta_T}^{\a ax{j+1}n}_{\a cyj{n-1}} \,.
\end{equation}

\noindent
The quantities $\SG$ have definite transversal properties

\begin{eqnarray}
\delta_T \cdot \SG &=& \delta_T \,,
\\
\SG \cdot \delta_T &=& \SG \,,  \label{sigmaT}
\end{eqnarray}

\noindent
and under a change of the prescription $\phi^{ax}_{1\,y} \to
\phi^{ax}_{2\,y}$ we have

\begin{equation}
\SG [\phi_1] = \SG [\phi_2] \cdot \SG [\phi_1]
\label{sigma21}
\end{equation}

\noindent
Let us note that (\ref{ESZ})  connects  the  component  $E^{\vc\mu
1n}$ of rank n with the
components of $\vec{Z}$ of rank one to n. That is

\begin{equation}
E^{\vc\mu 1n} = \sum^n_{m=1} \SG\ind{}{\vc\mu 1n}{\vc\nu 1m} \;
Z^{\vc\nu 1m}
\end{equation}

It is useful to introduce a vector product on ${\cal E}^*$. In general,
given
two vectors $\vec{E}_1$ and $\vec{E}_2$ we define its $\times$-product as
the vector $(\vec{E}_1 \times \vec{E}_2)$ with components

\begin{equation}
(\vec{E}_1  \times  \vec{E}_2)^{\vc\mu  1n}   =   \sum^{n-1}_{i=1}
E_1^{\vc\mu 1i}\,E_2^{\vc\mu{i+1}n}\;\;.
\label{contraccion}
\end{equation}

\noindent
The $\times$-product is associative and have the important property that
satisfies the differential constraint if any $\vec{E}_j$ does.

Equations (\ref{Z}) and (\ref{ESZ}) define an isomorphism between the
vector spaces ${\cal E}^*$ and ${\cal Z}$. The vector product can be
introduced in the vector space ${\cal Z}$ and, due to the isomorphism we
have

\begin{equation}
\vec{E}_1 \times \vec{E}_2 \;= \; \SG \; \cdot \;
(\vec{Z}_1 \times \vec{Z}_2) \;.
\end{equation}

It would be important for our purpose to consider the subset ${\cal F}
\subset
{\cal E}^*$ of elements that satisfy the homogeneous algebraic constraint.
Any element $\vec{F} \in {\cal F}$ identically vanish under the action of
the permutation operator associated with the algebraic constraint

\begin{equation}
F^{\vk \mu 1k{k+1}n} = 0 \;\; .
\end{equation}
This set is a subspace of ${\cal E}^*$. The isomorphism between the spaces
${\cal E}^*$ and ${\cal Z}$ induces a one to one correspondence between the
vectors of ${\cal F}$ and a subspace ${\cal Y} \subset {\cal Z}$. In the
appendix B we demonstrate that ${\cal Y}$ is the set of {\it all}
vectors of ${\cal Z}$ that satisfy the homogeneous algebraic
constraint.

\subsection{\bf Non-trivial representations of the diffeomorphism group}

Here, we study the transformation laws of the elements of the vector space
${\cal Z}$ under diffeomorphism transformations with a fixed point. The
vectors $\vec{E} \in {\cal E}^*$ behave as multivector densities, that is

\begin{equation}
\vec{E}' \; = \; \Lambda_D \; \cdot \; \vec{E} \;.
\end{equation}

\noindent
Thus

\begin{equation}
\vec{Z}' = \delta_T \; \cdot \; \Lambda_D \; \cdot \; \vec{E}
 = {\cal L} _D \cdot \; \vec{Z}
\end{equation}
with

\begin{equation}
{\cal L} _D \equiv \delta_T \; \cdot \; \Lambda_D \; \cdot \;
\SG \;.
\end{equation}

The isomorphism between the vector spaces ${\cal E}^*$ and ${\cal Z}$ makes
${\cal L} _D$ a representation of the diffeomorphism group. This
representation emerges as the push-forward of the natural action of
diffeomorphisms on the space of solutions of the differential constraints
by the isomorphism of that space with the space of transverse vectors
${\cal Z}$.

The presence of the nondiagonal matrix $\SG$ in ${\cal L} _D$ makes this
representation highly non trivial. This is an important result, due to the
possibility to introduce "spinor-like" objects in a theory invariant under
diffeomorphisms. In fact, the isomorphism guarantees the following property
of the $\SG$'s

\begin{equation}
\SG = \Lambda_D \cdot \SG \cdot {\cal L} _{D^{-1}}
\end{equation}
This relationship clearly shows the role played by the $\SG$'s as the
soldering quantities between the "vector" representation $\Lambda_D$ and
the "spinor-like" representation ${\cal L}_D$.

It is straightforward to see that the subspaces ${\cal F}$ and ${\cal Y}$
are invariants under diffeomorphisms. The vector product of $
\vec{Y}'s$ belongs to ${\cal Z}$ and in consequence has the same
transformation
law under diffeomorphisms that a single $\vec{Y}$.

\begin{equation}
\bigl( {\vec{Y}'}_1\,\times\, \cdots \,\times\,
{\vec{Y}'}_n \bigr) = {\cal L}_D \, \cdot \, \bigl( \vec{Y}_1
\,\times\,\cdots \,\times\, \vec{Y}_n\bigr) \, .
\label{variosY}
\end{equation}

\noindent
This fact will be important for the construction of knot invariants.

\section{\bf Free coordinates on the loop space}

We proceed now to determinate the free coordinates associated with an
arbitrary element of the loop space. This means that we have to
identify all those elements of the subspace ${\cal Y}$ of the space of
solutions of the homogeneous differential constraint that can be put in
correspondence with loops.

Let us start defining the algebraic-free coordinates $\vec{F}(L) \in {\cal
F}$ of the loop L (an  alternative  expression  will  be  given  in
section 7). A complete solution of the full set of algebraic
equations can be obtained in terms of the combinations

\begin{eqnarray}
\vec{R}(L) &=& \vec{X}(L) + \vec{X}(\overline {\vphantom{{}^1}L})
  \nonumber
\\ && \\
\vec{P}(L) &=& \vec{X}(L) - \vec{X}(\overline {\vphantom{{}^1}L})
 \nonumber
\end{eqnarray}

\noindent
We notice that $R^{\vc\mu 1n}$ is always reducible to a product of X's with
smaller rank. From the identities

\begin{eqnarray}
\vec{X}(L_1 \circ L_2) = &&\vec{X}(L_1) + \vec{X}(L_2) +
\vec{X}(L_1) \times \vec{X}(L_2)
\\ && \\
X^{\vc\mu 1n}(\overline {\vphantom{{}^1}L}) = &&(-1)^n X^{\vc\mu n1}(L)
\nonumber
\end{eqnarray}
and using the previous definitions one immediately obtains

\begin{equation}
\vec{R}(L) = \frac 14 \;(\vec{P}(L) \times \vec{P}(L))
- \frac 14 \;(\vec{R}(L) \times \vec{R}(L))
\end{equation}

\noindent
The last equation enables to express $\vec{R}(L)$ in terms of even products
of $\vec{P}$'s. We conclude that the algebraic-free coordinates are
contained in the combination $\vec{P}(L)$. Applying the first constraint on
$\vec{P}(L)$ one obtains

\begin{equation}
P^{\underline{\mu_1} \vc\mu 2n}(L) = X^{\mu_1}(L) R^{\vc\mu 2n}(L) \;\;.
\end{equation}

Both $\vec{P}(L)$ and $\vec{R}(L)$ satisfy the differential constraint and
have a definite behavior under the inversion of the orientation of the
loop,

\begin{equation}
\vec{R} (\overline {\vphantom{{}^1}L}) = + \vec{R}(L) \;\;,\phantom{PP}
\vec{P} (\overline {\vphantom{{}^1}L}) = - \vec{P}(L) \;\;.
\end{equation}

\noindent
We observe that the $\times$-product $(\vec{P} \times \vec{R}^m) \equiv
(\vec{P} \times \vec{R} \times \stackrel{m \, times}{\ldots} \times
\vec{R})$ has the symmetry of the $\vec{P}$'s under inversion of the loop and

\begin{equation}
(P \times R^m)^{\underline{\mu_1} \vc\mu 2n} = X^{\mu_1} \biggl[ 2(1+2m)
(R^m)^{\vc\mu 2n }+ (m+1) (R^{m+1})^{\vc\mu 2n} \biggr] \;.
\end{equation}

\noindent
Then the following combination

\begin{equation}
\vec{F}(L) = \vec{P}(L) + \sum_{m=1}^{\infty} \alpha_m (\vec{P}(L) \times
\vec{R}^m(L))
\label{Fm}
\end{equation}
satisfies the constraint $F^{\underline{\mu_1} \vc\mu 2n} = 0$ if the
coefficients take
the values

\begin{equation}
\alpha_m = \frac {(-1)^m m!}{2^m(2m+1)!!} \;\;.
\end{equation}

The action of a higher order algebraic constraint on $(P\times R^m)$ can be
computed in a similar  way.  It  is  found  that  the  combination
(\ref{Fm}) is
annihilated by the constraint for the same values of the coefficients
$\alpha_m$ in all cases. The algebraic-free coordinate $\vec{F}$ given by
(\ref{Fm}) automatically satisfy the differential constraint. This result
can be inverted to express $\vec{X}$ in terms of the coordinates $\vec{F}$
leading to an expansion containing the $\times$-product of any number of
$\vec{F}$'s.

All the $\vec{F}(L)$'s and its vector products belong to the space $
{\cal E}^*$. In consequence, one can generate new gauge covariant quantities,
others than (\ref{holonomia}), replacing in (\ref{holoE}) $\vec{E}$ by
$\vec{F}$ or the $\times$-product of any number of F's.

The free coordinates $\vec{Y}(L)$ are defined through the isomorphism
between the vector spaces ${\cal F}$ and ${\cal Y}$. We get

\begin{equation}
\vec{Y}(L) = \delta_T \cdot \vec{F}(L)
\end{equation}
and

\begin{equation}
\vec{F}(L) = \SG \cdot \vec{Y}(L)
\end{equation}

Due to the isomorphism, it is clear that we are using the term
"coordinate" of a loop in a wide sense. The one to one correspondence is
established between objects that belong to more general spaces that contain,
among others, all the elements associated with loops. In spite of this
fact, we choose to call $\vec{Y}(L)$ the coordinates of the loop $L$ since
this terminology seems to us the best way to describe the meaning of these
objects.

To conclude, the loop coordinates $\vec{Y}(L)$ have been identified as the
transverse part of the algebraic independent quantities $\vec{F}(L)$. Both
are related by a linear transformation and the coordinates $\vec{Y}$ of
certain loop L depend on the projection prescription.

\section{\bf Knot invariants}

In this section we analyze the relationship between knot invariants
and the invariant forms defined on the space ${\cal Y}$. Covariant vectors
are forms in the linear space ${\cal Y}$ such that

\begin{equation}
{\bf f} (\alpha \, \vec{Y}_1 + \beta \, \vec{Y}_2) \; = \; \alpha \,
{\bf f} ( \vec{Y}_1) + \beta \, {\bf f} ( \vec{Y}_2)
\end{equation}
where ${\bf f}$ is the row matrix ${\bf f} = ( f_{\mu_1}, \ldots , f_{\vc
\mu 1n}, \ldots )$. Covariant vectors are defined up to gradients. However a
particular transverse prescription can be fixed using

\begin{equation}
{\bf f} \; \equiv \; {\bf f} \; \cdot \; \delta_T \; .
\end{equation}
When a diffeomorphism transformation is performed, the transformed
covariant vectors will be in the same transverse prescription and are given
by

\begin{equation}
{\bf f}' \; = \; {\bf f} \; \cdot \; {\cal L}_{D^{-1}}
\end{equation}

Consider now the following quantity

\begin{equation}
{\cal I}(L) = {\bf g} \; \cdot \; \left( \vec{Y}(L) \times \stackrel{n \;
times}{\ldots} \times \vec{Y}(L) \right)
\end{equation}

\noindent
where ${\bf g}$ is a covariant vector and $\vec{Y}(L)$ gives the
coordinates of the loop L. If ${\bf g}$ is a covariant invariant vector,

\begin{equation}
{\bf g}' \; = \; {\bf g} \; \cdot \; {\cal L}_{D^{-1}} \; =
{\bf g} \;,
\end{equation}
from equation (\ref{variosY}) it is immediate to conclude that ${\cal
I}(L) = {\cal I}(L')$, being $L'$ the diffeomorphism transform of L. In
other words, ${\cal I}(L)$ is a knot invariant.It is important  to
remark that we are not going  to  discuss  analytical  details  of
regularization and consequently this statement is formal.

The first non trivial invariant tensor ${\bf g}_G$ has only one non
vanishing component

\begin{equation}
g_{G \, \vc\mu 1n} = \delta_{n,2} \, g_{\mu_1\mu_2}
\end{equation}

\noindent
where $g_{\mu_1\mu_2}$  is the previously obtained metric (\ref{g}), and
leads to the Gauss knot invariant

\begin{equation}
I_G = {\bf g}_{G} \cdot (\vec{Y} \, \times \, \vec{Y}) =
g_{\mu_1\mu_2} \, Y^{\mu_1}Y^{\mu_2} \, .
\end{equation}

The second non trivial invariant tensor is

\begin{equation}
g_{A \, \an ax} = \delta_{n,3} \, h_{\mu_1\mu_2\mu_3}
+ \delta_{n,4} \, g_{\mu_1\mu_3} \,g_{\mu_2\mu_4}
\end{equation}

\noindent
with

\begin{equation}
h_{a_1x_1a_2x_2a_3x_3} = \int d^3y \, \epsilon^{def}
\,g_{dya_1x_1} \,g_{eya_2x_2} \,g_{fya_3x_3}
\end{equation}

\noindent
and leads to the Alexander Conway $ 2^{ND} $ coefficient\cite{Guada}

\begin{eqnarray}
I_A &=& {\bf g}_{A} \cdot (\vec{Y} \, \times \, \vec{Y}) \nonumber
\\
&=& 2h_{\mu_1\mu_2\mu_3} Y^{\mu_1} Y^{\mu_2 \mu_3}
\\
&+&\,g_{\mu_1\mu_3} \,g_{\mu_2\mu_4} ( Y^{\mu_1 \mu_2}
Y^{\mu_3 \mu_4} + 2Y^{\mu_1 \mu_2\mu_3} Y^{\mu_4} ) \;.
\nonumber
\end{eqnarray}

Notice that one can directly express knot invariants in terms of the
the algebraic-free coordinates $\vec{F}$'s. In fact, ${\bf g} \; \cdot \;
(\vec{Y}^n) \; = \; {\bf g} \; \cdot \; (\vec{F}^n)$ due to the
transverse character of the invariant tensor ${\bf g}$ and the definition of
$\vec{Y}$.

Knot invariants are prescription independent. Let fix some prescription for
${\bf g}$, $\; {\bf g}_1 \; = \; {\bf g}_1 \; \cdot \; \delta_{T1}
\;$. Then

\begin{equation}
{\bf g}_1 \; \cdot \; \vec{Y}_1 \; = \; {\bf g}_1 \; \cdot \; \delta_{T1} \;
\cdot \; \vec{F} \; = \; {\bf g}_1 \; \cdot \; \vec{F}
\end{equation}

\noindent
But $\; \vec{F} \; = \; \SG_2 \; \cdot \; \vec{Y}_2 \; $, then

\begin{equation}
{\bf g}_1 \; \cdot \; \vec{Y}_1 \; = \; {\bf g}_1 \; \cdot \; \SG_2 \;
\cdot \; \vec{Y}_2 \; = \; {\bf g}_2 \; \cdot \; \vec{Y}_2
\end{equation}

\noindent
where

\begin{equation}
{\bf g}_2 \; = \; {\bf g}_1 \; \cdot \; \SG_2
\end{equation}
is the invariant tensor in the prescription 2. Using the algebraic-free
coordinates we have

\begin{equation}
{\bf g}_1 \; \cdot \; \vec{F} \; =
{\bf g}_1 \; \cdot \; \vec{Y}_1 \; =
{\bf g}_2 \; \cdot \; \vec{Y}_2 \; = {\bf g}_2 \; \cdot \; \vec{F}
\end{equation}

In Quantum Gravity one can use the Mandelstam identities for
SU(2)\cite{R14},

\begin{equation}
W_A(L_1)W_A(L_2) = W_A(L_1L_2) + W_A(L_1 \overline{\vphantom{ {}^1 }L}_2)
\; , \label{mandelstam}
\end{equation}
in order to give a general expression of the knot invariants involved in
the physical state space. Taking $L_1$ as the identity loop we find that
the Wilson loop functional have the symmetry of the R's under inversion of
the loop orientation. So in this case these functionals would depend only
on the cyclic permutation of $R^{\vc\mu 1n}$. Making use of eq. (\ref{psi})
the one loop states take the form

\begin{equation}
\Psi(L) = \sum_{n=1}^\infty \psi_{\vc\mu 1n}
R_c^{\vc\mu 1n}(L) \;.
\end{equation}

The cyclic permutations $R_c^{\vc\mu 1n}$ can be written in terms of even
vectorial products of $\vec{F}$'s or $\vec{Y}$'s coordinates, so

\begin{equation}
\Psi(L) = \sum_{m=1}^\infty \; \beta_m \; {\bf \psi} \; \cdot \;
(\vec{F}^{2m})_c  \;
= \sum_{m=1}^\infty \; \beta_m \; {\bf g} \; \cdot \;
(\vec{Y}^{2m})_c  \label{psiL}
\end{equation}

\noindent
where

\begin{equation}
{\bf g} \; \equiv \; {\bf \psi} \; \cdot \; \SG \;.
\end{equation}
The coefficients $\beta_i$ that appear in the expansion of $R_c$ in terms
of F can be directly evaluated using the results of section (3.1).

The wavefunction $\Psi$ would represent a quantum state of the
gravitational field if it is annihilated by all the constraints of
quantum gravity\cite{R14}. One of these constraints is the generator of
diffeomorphisms on a three dimensional manifold. Then, a {\it cyclic}
invariant tensor in the ${\cal Y}$ space and a quantum state of general
relativity turn out to be closely related. The cyclic property of the
invariant tensor means independence with respect to the base point taken as
the origin of the loops (the diffeomorphism transformations considered here
do not contain general translations). Equation (\ref{psiL}) shows the
general structure of such a state: it is a multilinear invariant form
constructed by even vectorial products of the vector $\vec{Y}$ with itself.
The restriction to even products is a direct consequence of the Mandelstam
identities for SU(2).

Let us make some remarks about this result. By (\ref{variosY}), each term
of the sum is a knot invariant in its own right. But the invariant tensor
involved in the whole sum is the same, so a {\it family} of knot invariants
is associated with each invariant tensor ${\bf g}$. In the above examples,
${\bf g}_G$ gives only the Gauss invariant, but ${\bf g}_A$ produces also
other knot invariant when is contracted with $(\vec{Y}^4)$. In this case it
reduces simply to $(I_G)^2$. In general, an invariant tensor involving at
most rank 2p components allows to construct p in principle independent knot
invariants. This fact is a direct consequence of the transformation
properties of the $\times$-product of contravariant transverse objects.

To conclude this section, we show another interesting result related with
the construction of link invariants from knot invariants in the case of
Quantum Gravity. Let $\Psi(L)$ be a knot invariant of Quantum Gravity and
let $L_1 \; L_2$ two arbitrary loops. Then consider any open path P
connecting the origins of $L_1$ and $L_2$. From (\ref{psi}) and the
Mandelstam identity (\ref{mandelstam}) we know that the combination
$\Psi(L_1 \; P \; L_2 \; \overline {\vphantom{{}^1}P}) + \Psi(L_1 \; P \;
\overline {\vphantom{{}^1}L}_2 \; \overline {\vphantom{{}^1}P})$ is
independent of P and it only depends on the original loops $L_1$ and $L_2$.
This means that

\begin{equation}
\Psi(L_1, L_2) \; = \; \Psi(L_1 \; P \; L_2 \; \overline {\vphantom{{}^1}P})
+ \Psi(L_1 \; P \; \overline {\vphantom{{}^1}L}_2 \; \overline
{\vphantom{{}^1}P})
\end{equation}
is a link invariant.

\section{\bf The Extended Loop Group}

In the previous sections it was shown  that any element of the loop space
can be described by a set of multivector density fields constrained by a
set of algebraic and differential identities. We develop now the basis of a
more complete and rigorous mathematical description of these objects.

We begin by considering the set of all objects of the type

\begin{equation}
{\bf E} \; = \; (\, E, \, E^{\mu_1}, \, \cdots , \, E^{\vc\mu 1n},
\, \cdots ) \; \equiv (\, E, \, \vec{E})
\end{equation}
where $E$ is a real number and $E^{\vc\mu 1n}$ (for any $n \neq 0$) is an
arbitrary multivector density field. This set has the structure of a
vector space (denoted as $\cal E$) with the usual composition laws of
addition and multiplication between functions.

A product law on $\cal E$ can be introduced as follows: given two vectors
${\bf E}_1$ and ${\bf E}_2$, we define ${\bf E}_1 \times {\bf E}_2$ as
the vector with components

\begin{equation}
{\bf E}_1 \times {\bf E}_2 \; = \; (\,E_1 E_2, \, E_1 \vec{E_2} \, +
\, \vec{E_1}
E_2 \, + \, \vec{E_1} \times \vec{E_2})
\label{producto}
\end{equation}
where  $\vec{E_1}  \times  \vec{E_2}$   is   given   by   equation
(\ref{contraccion}). The composition law (\ref{producto}) is an
extension of the vector
product among loop coordinates introduced in section 3.1.
For any value of n, the rank n component of the $\times$-product can be
expressed as

\begin{equation}
({\bf E}_1 \times {\bf E}_2)^{\vc\mu 1n}  =
\sum^{n}_{i=0} E_1^{\vc\mu 1i}\,E_2^{\vc\mu{i+1}n}
\end{equation}
with the convention
\begin{equation}
E^{\mu_1 \ldots \mu_0} = E^{\mu_{n+1} \ldots \mu_n} = E \;\; .
\label{convencion}
\end{equation}

The product law is associative and distributive with respect to the
addition of vectors. It has a null element (the null vector) and a identity
element, given by

\begin{equation}
{\bf I} = (\, 1,\, 0,\, \cdots ,\, 0, \, \cdots) \;\; .
\end{equation}

An inverse element exits for all vectors with nonvanishing zero rank
component. It is given by

\begin{equation}
{\bf E}^{-1} \; = \; E^{-1} {\bf I} \; + \; \sum^{\infty}_{i=1}
(-1)^i E^{-i-1}
({\bf E} \, - \, E {\bf I})^i \label{inverse}
\end{equation}
such that

\begin{equation}
{\bf E} \times {\bf E}^{-1} \; = \; {\bf E}^{-1} \times {\bf E}
 \; = \; {\bf I} \;\; .
\end{equation}
It should be noticed that, when evaluating the components of ${\bf
E}^{-1}$, the sum involved in (\ref{inverse}) is actually finite due to
the fact that $({\bf E} \, - \, E {\bf I})$ is a vector with its zero rank
component equal to zero. In consequence

\begin{equation}
\left[ \; ({\bf E} \, - \, E {\bf I})^i \; \right]^{\vc\mu 1n} \; = \; 0
\;\; {\em if} \;\; n < i \;\; .
\label{corte}
\end{equation}

The set of all vectors with nonvanishing zero rank component forms a group
with the $\times$-product law.

Now we concentrate on the set $\cal X$ of vectors of $\cal E$
that have its zero rank component equal to one, ${\bf X}\;=\;(1,\,
\vec{X})$, with the components of $\vec X$ multivector density fields
that obey the algebraic constraint (\ref{VA}) and differential constraint
(\ref{dc}).

The set $\cal X$ is closed under the $\times$-product law. If ${\bf X}_1
\in \cal X$ and ${\bf X}_2 \in \cal X$, it is clear from the definition of
the group product that ${\bf X}_1 \times {\bf X}_2$ satisfy the
differential constraint. One can also demonstrate  that ${\bf X}_1 \times
{\bf X}_2$ satisfies the algebraic constraint. In a similar way one can
show that the inverse ${\bf X}^{-1}$ given by (\ref{inverse}) satisfies the
constraints if $\bf X$ does. A detailed proof of these properties is given
in appendix C. This results show that the algebraic and differential
constraint preserves the group structure under the $\times$-composition
law. We call $\cal X$ the Special-extended Loop group (SeL group)
\cite{aclaracion}.

The group of loops is a subgroup of the SeL group since ${\bf X}(L) \; \in
\; \cal X$ and

\begin{equation}
{\bf X}(L_1 \circ L_2) = {\bf X}(L_1) \times {\bf X}(L_2)
\end{equation}
where $\circ$ indicates the group composition law in loop space.

We now show that this group is in fact larger than the group of loops. For
that, consider the group element ${\bf X}^m \, \equiv \, {\bf X} \times
\stackrel{\rm m}{\ldots} \times {\bf X}$. Note that if ${\bf X}$ gives the
multitangent field of certain loop L, ${\bf X}^m$ would be the multitangent
field of the loop L swept itself m times. We get by the binomial expansion

\begin{equation}
{\bf X}^m \equiv \left[ {\bf I} \, + \, ({\bf X} \, - \,
{\bf I}) \right]^m =  {\bf I} \, + \, \sum^{m}_{i=1} \c mi
({\bf X} \, - \, {\bf I})^i \;\; .
\label{potencia}
\end{equation}
The extension of (\ref{potencia}) to real values of m is straightforward,
being defined as

\begin{equation}
{\bf X}^\lambda = {\bf I} \, + \, \sum^{\infty}_{i=1}
\c \lambda i ({\bf X} \, - \, {\bf I})^i
\label{extension}
\end{equation}
with $\lambda$ real. We usually call this object the analytic extension of
$\bf X$. Note that for $\lambda \, = \, -1$ we recover the expression of
the inverse of $\bf X$. Also in this case, due to (\ref{corte}) the
analytic extension is well defined for all elements of $\cal X$. One can
prove that if $\bf X$ is constrained by the differential and algebraic
identities, its analytic extension also verifies the constraints (see
appendix C). So, the analytic extension of any $\bf X$ is in $\cal X$.
Moreover, we have

\begin{equation}
{\bf X}^\lambda \, \times \, {\bf X}^\mu \; = \; {\bf X}^{\lambda +
\mu} \;\; .
\end{equation}
We conclude that the set $\{ {\bf X}^\lambda\;/ \lambda\in \mbox{R and}
\,\; {\bf X}\in\cal X \}$ defines an abelian
one-parameter subgroup of the $\cal X$ group.

For non-integer values of $\lambda$, the analytic extension of a loop
coordinate is not a loop coordinate. This fact explicitly shows that there
exists in $\cal X$ others elements besides the loop coordinates.

Matrix representations of the SeL group can be generated through a natural
extension of the holonomy. The extended holonomy associated with a
nonabelian connection $A_{ax}$ is defined as $U_A({\bf X}) = \bf {A \cdot
\bf X}$,
where ${\bf A} \equiv ( 1, A_{a_1x_1},\cdots,A_{a_1x_1}\ldots
A_{a_nx_n},\cdots)$, ${\bf X} \equiv (1, X^{a_1x_1}, \cdots, X^{\an
ax},\cdots)$ and the dot acts like a generalized Einstein convention
including contractions of the discrete indices $a_i$ and the continuous
indices $x_i$.
. We have

\begin{eqnarray}
U_A({\bf X}_1) \, U_A({\bf X}_2) &=& \sum^{\infty}_{i=0} \,
\sum^{\infty}_{j=i} \, A_{\mu_1 \ldots \mu_i} A_{\mu_{i+1} \ldots \mu_j}
X_1^{\vc \mu 1i} X_2^{\vc \mu {i+1}j} \nonumber \\
&=& \sum^{\infty}_{j=0} \, A_{\mu_1 \ldots \mu_j}
\left( \sum^{j}_{i=0} \, X_1^{\vc \mu 1i} X_2^{\vc \mu {i+1}j} \right)
\,=\, U_A({\bf X}_1 \times {\bf X}_2)
\rule{0mm}{6mm}
\end{eqnarray}
where the convention (\ref{convencion}) was applied over all the
indices. The correspondence ${\bf X} \rightarrow U_A({\bf X})$ gives
a representation of the SeL group into a particular gauge group.
The differential constraint imposed on $\bf X$ assures $U_A({\bf X})$ to be
a gauge covariant quantity.

We have shown that the analytic extension of any element of the SeL group
defines a one-parameter subgroup. By studying its properties one can find
the algebra associated with the SeL group.

\section{\bf The algebra of the SeL group}

\subsection{\bf Generators}

Consider the one-parameter subgroup $\{ {\bf X}^\lambda \}$ and
suppose that $\lambda$ varies in an infinitesimal amount. We can write

\begin{equation}
{\bf X}^{\lambda \, + \, d \lambda} \; = \;
{\bf X}^\lambda \, \times \, {\bf X}^{d \lambda} \; = \,
{\bf X}^\lambda \, + \, \frac{d{\bf X}^\lambda}{d \lambda} \, d \lambda
\;\;.
 \label{uno}
\end{equation}
Taking $\lambda = 0$

\begin{equation}
{\bf X}^{d \lambda} \; = \; {\bf I} \; + \; {\it F} \, d \lambda
 \label{dos}
\end{equation}
where

\begin{equation}
{\it F} \; \stackrel{\rm def}{=} \; \frac{d{\bf X}^\lambda}{d \lambda}
\mid_{\lambda=0} \; = \; ( \, 0, \, \sum^{\infty}_{i=1} \;
\frac{(-1)^{i-1}}{i} \;
\vec{X}^i) \; = \;(\, 0, \, \vec{F}) \;\;.
 \label{F}
\end{equation}
Introducing (\ref{dos}) in (\ref{uno}) we obtain the following differential
equation for the elements of $\{ {\bf X}^\lambda \}$

\begin{equation}
\frac{d{\bf X}^\lambda}{d \lambda} \; = \;{\bf X}^\lambda \, \times \, {\it
F} \; = \; {\it F} \, \times \, {\bf X}^{\lambda} \;\;.
\end{equation}
This equation can be integrated, obtaining

\begin{equation}
\rule{0mm}{10mm}
{\bf X}^\lambda = {\bf I} + \sum^{n}_{k=1} \, \frac{\lambda^k}{k!}{\it
F}^k + {\it F}^{n+1} \times \int_{0}^{\lambda}d \lambda_1
\int_{0}^{\lambda_1}d \lambda_2 \cdots \int_{0}^{\lambda_n}d \lambda_{n+1}
\, {\bf X}^{\lambda_{n+1}} \;\;.
\end{equation}
The iterative integration actually stops for any finite rank n component
(${\it F}^{n+1} = {\it F} \times \stackrel{n+1}{\dots} \times {\it F}= 0$
in this case). So

\begin{equation}
\rule{0mm}{10mm}
{\bf X}^\lambda = {\bf I} + \sum^{\infty}_{k=1} \, \frac{\lambda^k}{k!}{\it
F}^k = \exp (\lambda {\it F}) \;\;. \label{exp}
\end{equation}
We conclude that the vector ${\it F}$ given by (\ref{F}) is
the generator of the one-parameter subgroup $\{ {\bf X}^\lambda \}$. It is
evident that the generator satisfy the differential constraint. We prove
now the following fundamental property: ${\it F}$ satisfies the homogeneous
algebraic constraint. In other words, the generator of the one-parameter
subgroup $\{ {\bf X}^\lambda \}$ is the algebraic free part of $\bf X$.

We know that

\begin{equation}
( \,{\bf X}^\lambda \, )^{\vk \mu 1k{k+1}n} =
( \,{\bf X}^\lambda \, )^{\vc \mu 1k} \,
( \,{\bf X}^\lambda \, )^{\vc \mu {k+1}n} \;\;.
\end{equation}
Differentiating with respect to $\lambda$ and evaluating for $\lambda = 0$
we get

\begin{equation}
\frac {d}{d \lambda} \left( \,{\bf X}^\lambda \, \right)^{\vk \mu
1k{k+1}n}_{\lambda =
0} = \left( \, \frac{d {\bf X}^\lambda}{d \lambda} \, \right)^{\vc \mu
1k}_{\lambda =
0}{\bf I}^{\vc \mu {k+1}n} + {\bf I}^{\vc \mu 1k}
\left( \, \frac{d {\bf X}^\lambda}{d \lambda} \, \right)^{\vc \mu
{k+1}n}_{\lambda =
0} \;.
\end{equation}
As $1 \leq k < n$, we conclude

\begin{equation}
{\it F}^{\vk \mu 1k{k+1}n} \, = \, 0 \;\;,\;\; 1 \leq k < n \;\;.
\end{equation}
Reciprocally, one can demonstrate that the exponential of any algebraic-free
quantity produces an object that satisfy the algebraic constraint. It is
important to stress that these results permit to obtain a general solution
for the algebraic constraint (equation (\ref{exp}) with $\lambda =
1$ and its inverse (\ref{F}) gives the relationship between an object that
satisfy the algebraic constraint and its algebraic-free part).

\subsection{\bf The algebra}

The set of all $\it F$'s that satisfies the differential constraint and
are annihilated by the algebraic constraint forms a  vector  space
$\cal F$ \cite{aclare}. One can define a bilinear operation on
$\cal F$ in the following way,

\begin{equation}
\left[ \it{F}_1 ,\it{F}_2 \right] = \it{F}_1 \times \it{F}_2 -
\it{F}_2 \times \it{F}_1 \;\;\; {\rm for \; any}  \; \; \it{F}_1 ,\it{F}_2
\in \cal F.
\label{conmutador}
\end{equation}
This operation is closed on $\cal F$. The vector space $\cal F$ together
with the bracket operation (\ref{conmutador}) defines the algebra
associated to the SeL group.

A basis of this algebra can be found by means of the transverse part of the
algebraic free quantities. The longitudinal and transverse orthogonal
projectors acting on the space of transverse vector densities were
defined in section 3.1.

The transverse part of any vector ${\it F} \in {\cal F}$ is then given by

\begin{equation}
{\bf Y} = \delta_T \cdot {\it F}
\label{coor}
\end{equation}
When ${\it F} = {\it F}(L)$, equation (\ref{coor}) gives the free
coordinates of the loop L defined  in  section  3.2.  The  inverse
relation is

\begin{equation}
\it F \;= \; \SIGMA \; \cdot \; \bf Y
\label{Fdos}
\end{equation}
where   $\SIGMA$   are   the   soldering   quantities.

The homogeneous algebraic constraint can be imposed by defining a
projector $\OMEGA$ over the algebraic free part of an arbitrary vector.
Consider the matrix
\begin{equation}
\delta\ind{}{\vc \mu 1n}{\vc \nu 1m} = \delta_{n,m} \delta^{\mu_1}_{\nu_1}
\cdots \delta^{\mu_n}_{\nu_n}
\end{equation}
and the vector

\begin{equation}
{\bf\delta}_{\vc \nu 1i} \,=\, (\;0,\;
\delta\ind{}{\mu_1}{\vc \nu 1i}, \; \cdots ,
\delta\ind{}{\vc \mu 1n}{\vc \nu 1i}, \; \cdots
\,) \;\;.
\end{equation}
We define then a new matrix, given by

\begin{eqnarray}
\OMEGA\ind{}{\vc \mu 1n}{\vc \nu 1m} &\equiv& \frac{\delta_{n,m}}{m}
[\,[ \, \cdots \, [ \delta_{\nu_1} ,\delta_{\nu_2} ] , \, \cdots \, ],
\delta_{\nu_n}]^{\vc \mu 1n} \theta(m-1) \, + \, \delta_{m,1}
\delta^{\mu_1}_{\nu_1}
\nonumber \\ \label{omega} \\
&=& \delta\ind{}{\vc \mu 1n}{\vc \nu 1m} +
\sum^{n-1}_{k=1} \, \frac{(n-k)}{n} \, (-1)^k \,
\delta\ind{}{\vk \mu k1{k+1}n}{\vc \nu 1m} \nonumber
\end{eqnarray}
The matrix $\OMEGA$ has the following important property: it satisfies the
homogeneous algebraic constraint in the upper indexes. This fact
immediately shows that $\OMEGA$ is a projector. Given an arbitrary
vector $\bf E$, $\; \OMEGA \cdot \bf E$ is an algebraic-free object. In
particular we have ${\it F}=\OMEGA\cdot{\it F}$.

Let us introduce now the following set of vectors
\begin{equation}
{\base}_{\vc \nu 1m} \, = \, (\, 0, \, \vec{\base}_{\vc \nu 1m})
 \rule{0mm}{6mm}
\end{equation}
with
\begin{equation}
{\base}^{\vc \mu 1n}_{\vc \nu 1m} = \left( \SIGMA \cdot \OMEGA \right)
\ind{}{\vc \mu 1n}{\vc \nu 1m} \;\;. \rule{0mm}{6mm}
\end{equation}

The vectors $\base_{\vc \nu 1m}$ belong to $\cal F$. They form a linear
independent set of vectors that generates all the vector space $\cal F$.
 From (\ref{Fdos}) one can write

\begin{equation}
{\it F} \;= \; \base \; \cdot \; {\bf Y} \;\;.
\end{equation}
The set $\{ \base_{\vc \nu 1m} \}$ defines then a base of the algebra. The
components of any element of the algebra are simply given in this base by
the transverse part of the vector.

In order to obtain the structure constants associated with
$\{ {\base}_{\vc \nu 1m} \}$ one can start evaluating the commutator of two
$\OMEGA$'s. One finds
\begin{equation}
\left[ \OMEGA_{\vc \nu 1n} , \OMEGA_{\vc \mu 1m} \right]
= \frac{n+m}{2} \OMEGA \cdot \left(\frac{1}{n} \delta_{\vc \nu 1n}
\times\OMEGA_{\vc \mu 1m} -
\frac{1}{m} \delta_{\vc \mu 1m} \times \OMEGA_{\vc \nu 1n} \right)
\end{equation}
This relation can be obtained by first evaluating $\left[ \OMEGA_{\vc \nu
1{n-1}} , \delta_{\nu_n} \right]$ and then proceeding
by induction. Now it is straightforward to calculate the commutator of
two elements of the base. We obtain

\begin{equation}
\left[ {\base}_{\vc \nu 1n} , {\base}_{\vc \mu 1m} \right] =
\base \cdot f_{\vc \nu 1n,\vc \mu 1m}
\end{equation}
where the structure constants $f^{\vc\rho 1k}_{\vc \nu 1n,\vc \mu 1m}$ are
given by

\begin{equation}
f^{\vc\rho 1k}_{\vc \nu 1n,\vc \mu 1m} = \left[
\delta_T\cdot\OMEGA_{\vc \nu 1n},\delta_T\cdot\OMEGA_{\vc \mu 1m}
\right]^{\vc\rho 1k} \;\;. \label{sc}
\end{equation}

\subsection{\bf Diffeomorphism Invariants}

Any vector ${\it F}$ belonging to the SeL algebra is a multivector
density field under a  general  diffeomorphism  transformation  In
matrix form the transformation law is

\begin{equation}
{\it F}' = \Lambda_D \; \cdot \; {\it F}
\label{transfF}
\end{equation}
while the
transformation law of the transverse algebraic-free vectors ${\bf
Y}$ was derived in section 3.2

\begin{equation}
{\bf Y}' = \delta_T \cdot {\it F}' = \LL_D\cdot{\bf Y}
\end{equation}
where
\begin{equation}
\LL_D \equiv \delta_T \; \cdot \; \Lambda_D \; \bf\cdot \;
\SIGMA \;\;.
\end{equation}

The general diffeomorphism transformation considered in (\ref{transfF}) is
in fact a particular example of an automorphism of the algebra. Another
automorphism transformations can be considered, for example those induced by
the conjugation classes ${\bf X} \times {\it F} \times {\bf X}^{-1}$.

Diffeomorphism invariants can be related with some kind of invariant
forms defined in
the vector space ${\cal F}$. Consider a covariant vector ${\bf g} = (0, \,
g_{\mu_1 \mu_2},\cdots, \, g_{\vc \mu 1n},\cdots)$ that satisfies the
following properties:

\begin{eqnarray}
{\bf g} & = & {\bf g} \, \cdot \, \LL_D \label{auto1} \\
g_{\vc \mu 1n} \, & = & \, g_{(\vc \mu 1n)_{cyclic}} \;\; .
\label{auto2}
\end{eqnarray}
This tensor allows to define the following multilinear form

\begin{equation}
({\it F}_1,\cdots,{\it F}_n)_g =
{\bf g}\cdot ({\bf Y}_1\times\cdots\times {\bf Y}_n) \; \; ,
\label{multi}
\end{equation}
that is invariant with respect to both automorphism transformations just
considered. The invariance property (\ref{auto1}) together with Eq
(\ref{variosY}) makes (\ref{multi}) invariant with respect
diffeomorphism transformations with a fixed point. The ciclicity
property (\ref{auto2}) ensures invariance
with respect to the conjugation classes. Knot invariants are
formally generated evaluating on loops the
multilinear forms associated with the invariant tensors.

Each invariant tensor ${\bf g}$ is related with a {\it metric} function in
the algebra in the following way

\begin{equation}
({\base}_{\vc \nu 1n},{\base}_{\vc \mu 1m})_g =
\tilde{g}_{\vc \nu 1n\;,\vc \mu 1m} =
{\bf g}\cdot (\OMEGA_{\vc \nu 1n} \times
\OMEGA_{\vc \mu 1m}) \;\;.
\end{equation}

The metrics associated with the Gauss and Alexander-Conway invariants are
singular in the sense that there exist non null vectors with zero norm.
Several open questions about the general structure of the invariant forms
and its relationship with the irreducible representations of the algebra
remains to be studied. A complete study of the SeL group algebra is a non
trivial task since we are dealing with an infinite dimensional (Lie)
algebra.
\section{\bf Conclusions}

A coordinate representation on loop space has been introduced. In this
representation any loop is described by a set of contravariant tensors of
any rank defining a vector in the space of the transverse objects
${\cal Y}$. The loop
coordinates depend on a prescription due the
non uniqueness of the transverse part of a tensor. However knot invariants
turns out to be independent of the prescription, as it should be. Each
invariant form defined on ${\cal Y}$ generates a family of knot invariants.
The loop coordinates provide the basis to define an affine geometry
supported on the loop space.

 Within this approach we have shown
how the group of loops can be embedded in a natural way in a more general
algebraic structure, the Special extended Loop group. The
existence in the SeL group of other
elements besides the multitangent fields associated with loops has been
explicitly demonstrated. The algebra of the SeL group has been studied and
a primary approach to the formulation of invariant forms in the vector
space of the algebraic free coordinates has been done.

It is important to remark the connection between the solution of the
constraints and the new emerging group structure. The elements of the
algebra are algebraic-free vectors that satisfy the differential
constraint. The free-coordinates $\bf Y$ may be considered as free
parameters of the group. The soldering quantities
$\SIGMA$ connects two representations of the diffeomorphism group. One can
prove in a direct way that the following property holds, $\Lambda_D \cdot
\SIGMA \cdot \LL_D = \SIGMA$. So, the matrix $\SIGMA$ connects the "vector"
representation $\Lambda_D$ with the "spinor-like" representation $\LL_D$.
They act then like the Pauli matrices in an infinite dimensional Lie
algebra.

The quantum formulation of General Relativity in the Extended Loop
Group becomes closer to the familiar configuration  space  of  any
quantum field theory.  The  extension  of  the  Hilbert  space  of
quantum gravity including smooth functions gives a new perspective
on the framing problem associated with knot invariants and in  the
search of a measure allowing to define an inner product on it.  In
the  cases  were   an   inner   product   is   known,   the   U(1)
case\cite{R33,Nori} and linearized gravitation\cite{ARS91}, it has
been defined by means of the loop coordinates. This fact  strongly
suggest  that  the  extended  loop  group  could  be  the  natural
framework of a "loop representation".

Finally, we would like to remember that it was the study of simultaneous
partial differential equations that led Lie to investigate continuous
transformation groups. From this point of view, it is not surprising that
many of the differential equations of mathematical physics can be solved
using the Lie group techniques. It becomes natural to think then that the
existence of the local infinite dimensional Lie group associated with loop
space would allow to treat functional problems related with gauge theories
and gravitation in a simpler form.

\section*{\bf Acknowledgments}

R. Gambini wishes to thank Abhay Ashtekar, Lee Smolin, Jorge Pullin and
Bernd Br\"ugmann  for fruitful discussions.

\newpage
\section*{\bf Appendix A}

In this appendix we list several useful  formulae  concerning  the
$\SG$ introduced in section 3.1. This quantities satisfy
several useful relations: their transversal properties are given
by

\begin{eqnarray}
\delta_T \cdot \SG &=& \delta_T \,, \label{A5}
\\
\SG \cdot \delta_T &=& \SG \,,  \label{A6}
\end{eqnarray}

\noindent
and they obey the differential constraint

\begin{eqnarray}
&&\frac {\partial \phantom{iii}} {\partial x_i^{a_i}} \,
\SG\ind{}{\a ax1i\,\ldots\,a_nx_n}{\vc\nu 1m} =
\nonumber \\ && \nonumber \\
&&\bigl( \,\delta(x_i-x_{i-1}) - \delta(x_i-x_{i+1}) \,\bigr)
\SG\ind{}{\a ax1{i-1}\,\a ax{i+1}n}{\vc\nu 1m} \;.
\label{A7}
\end{eqnarray}

\noindent
We show now that the vector product of two $\SG$'s reproduce another
$\SG$. From (\ref{A7}) we see that the $\times$-product of two $\SG$'s
constructed with its contravariant indices automatically satisfy the
differential constraint, so we can write

\begin{eqnarray}
&&(\SG_{\vc\nu 1j} \, \times \; \SG_{\vc\rho 1m})^{\vc\mu 1n} =
\nonumber \\
&&\sum^{\infty}_{k=1} \,\SG\ind{}{\vc\mu 1n}{\vc\alpha 1k}
(\SG_{\vc\nu 1j} \, \times \;\SG_{\vc\rho 1m})^{\vc\alpha 1k} \,.
\end{eqnarray}

\noindent
Then using (\ref{A5}) and (\ref{A6}) we obtain

\begin{eqnarray}
&&(\SG_{\vc\nu 1j} \, \times \; \SG_{\vc\rho 1m})^{\vc\mu 1n}
\rule{0mm}{6mm} \nonumber \\
&&= \sum^{\infty}_{k=1} \, (\SG \; \cdot \; \delta_T)\ind{}{\vc\mu 1n}
{\vc\alpha 1k}(\SG_{\vc\nu 1j} \, \times \;\SG_{\vc\rho 1m})^{\vc\alpha 1k}
\nonumber \\
&&= \sum^{\infty}_{k=1} \, \SG\ind{}{\vc\mu 1n}{\vc\alpha 1k}
( \; (\delta_T \; \cdot \; \SG)_{\vc\nu 1j} \, \times \;
     (\delta_T \; \cdot \; \SG)_{\vc\rho 1m} \; )^{\vc\alpha 1k}
\nonumber \\
&&= \sum^{\infty}_{k=1} \, \SG\ind{}{\vc\mu 1n}{\vc\alpha 1k}
( \delta_{T \; \vc\nu 1j} \, \times \;
  \delta_{T \; \vc\rho 1m} )^{\vc\alpha 1k}
\nonumber \\
&&= \sum^{\infty}_{k=1} \, \SG\ind{}{\vc\mu 1n}{\vc\alpha 1k}
 \delta\ind{T}{\vc\alpha 1k}{\vc\nu 1j \; \vc\rho 1m}
\end{eqnarray}

\noindent
that is to say

\begin{equation}
\SG\ind{}{\vc\mu 1n}{\vc\nu 1j \, \vc\rho 1m} =
(\SG_{\vc\nu 1j} \, \times \;\SG_{\vc\rho 1m})^{\vc\mu 1n} \,.
\end{equation}

\noindent
Under a change of the projection prescription we have

\begin{equation}
\phi^{ax}_{1\,z} \to \phi^{ax}_{2\,z} \,,\;\; \delta_{1\;T} \to \delta_{2\;T}
\,,\;\; \delta_{1\;L} \to \delta_{2\;L} \,,\;\; \SG_1 \to \SG_2
\end{equation}

\noindent
then the following relationships can be easily proved

\begin{eqnarray}
\delta_{1\;T} &=& \delta_{2\;T} \cdot \delta_{1\;T}
\\
\delta_{2\;L} &=& \delta_{2\;L} \cdot \delta_{1\;L}
\\
\SG_1 &=& \SG_2 \cdot \SG_1
\end{eqnarray}

Now we give the transformation laws of $\SG$ under a general diffeomorphism
transformation $x^a \to x'^a = D^a(x)$. We start by introducing the
quantities $\delta_{D\;T}$ and $\delta_{D\;L}$ defined by

\begin{equation}
\delta_{D\;T} \equiv \Lambda_{D^{-1}} \cdot \delta_{T} \cdot \Lambda_D
\;, \qquad
\delta_{D\;L} \equiv \delta - \delta_{D\;T} =
\Lambda_{D^{-1}} \cdot \delta_{L} \cdot \Lambda_D
\end{equation}

\noindent
where $\Lambda_D$ is given by (\ref{lD}). Using the identity

\begin{equation}
\frac {\partial \;\;} {\partial x^a} \Lambda\ind{D^{-1}}{ax}{by} =
- \frac {\partial \;\;} {\partial y^b} \delta(x-D^{-1}(y))
\end{equation}

\noindent
it may be immediately proved that $\delta_{D\;T}$ and $\delta_{D\;L}$ are
transverse and longitudinal projectors. The function $\phi_D$, which
characterize the projection prescription, is related with $\phi$ by

\begin{equation}
\phi\ind{D}{ax}{y} \; = \; J(x) \; \frac {\partial x^a}
{\partial D^b (x)} \; \phi_{ \phantom{b \, D(x)} D(y)}^{b \, D(x)} \;.
\end{equation}

\noindent
Making use of the identity

\begin{equation}
\Lambda\ind{D}{ax}{cz} \, \delta(t-z) \, \Lambda\ind{D^{-1}}{cz}{by} =
 \delta(t-D^{-1}(x)) \, \delta\ind{}{ax}{by}
\end{equation}

\noindent
we get

\begin{equation}
\SG_{D} \equiv \SG\left[ \phi\ind{D}{ax}{y} \right] =
\Lambda_{D^{-1}} \cdot \SG \cdot \Lambda_D \;,
\label{sigmaD}
\end{equation}

\noindent
where $\SG_{D}$ is the soldering quantity constructed with $\phi_D$.

\section*{\bf Appendix B}

We shall demonstrate in this appendix that there exist an isomorphism
between the vector space ${\cal F}$ of vectors that
satisfy the differential constraint and the homogeneous algebraic
constraint, and the space ${\cal Y}$ of vectors that satisfy both
homogeneous constraints.

Consider a vector $\vec{ F} \in {\cal F}$. It is straightforward to see
that $\delta_T \cdot \vec{ F}$ satisfies the homogeneous differential and
algebraic constraint. In consequence, $\delta_T \cdot \vec{ F} \in {\cal
Y}$.

The reciprocal states that any algebraic-free transverse vector must
generate when contracting with a $\SG$ an object that satisfy the
homogeneous algebraic identities.

Let us first consider an arbitrary vector $\vec{ E}$ that satisfies the
differential constraint, i.e. $\vec{ E}  =  \SG  \cdot  \vec{  Z}$
where $\vec{
Z}$ is the transverse part of the vector. It is straightforward to
show that

\begin{eqnarray}
\lefteqn{\frac{\partial}{\partial x^{a_i}_i}E^{\vk \mu 1k{k+1}n} =
\left[\, \delta(x_i - b_{ik}) - \delta(x_i - c_{ik}) \,\right]}
 \nonumber \\
& &\left\{ E^{\underline{\raya \mu 1ik}\,\vc \mu{k+1}n} \theta(k+1-i)
 +\, E^{\underline{\vc \mu 1k}\,\raya \mu{k+1}in}\theta(i-k) \right\}
\label{vavd}
\end{eqnarray}
where

\begin{equation}
b_{ik} = \left\{ \begin{array}{ll}
x_0 & \; \mbox{if $i=1$ or $i=k+1$} \\
x_{i-1} & \; \mbox{elsewhere}
\end{array} \right.
\;\; , \;\;
c_{ik} = \left\{ \begin{array}{ll}
x_0 & \; \mbox{if $i=n$ or $i=k$} \\
x_{i+1} & \; \mbox{elsewhere}
\end{array} \right.
\end{equation}
For $n = 2$ we obtain from the last equation

\begin{equation}
\frac{\partial}{\partial x^{a_i}_i}E^{\underline{\mu_1} \mu_2} = 0
\end{equation}
so $E^{\underline{\mu_1} \mu_2}$ is a transverse vector. Suppose now
that $\vec{ Z} \in {\cal Y}$, that is

\begin{equation}
\vec{ Z}^{\vk \mu 1k{k+1}n} = 0 \;\;\;\; 1 \leq k <n
\end{equation}
For $n=2$ we can conclude that $E^{\underline{\mu_1} \mu_2} =
Z^{\underline{\mu_1} \mu_2} = 0$. We now proceed by induction. Suppose
that

\begin{equation}
\vec{ E}^{\vk \mu 1k{k+1}n} = 0 \;\;\;\;\; \forall n \in [2,h] \; ,
\;\forall k \in [1,n-1]
\end{equation}
For $n = h+1$, (\ref{vavd}) shows that $E^{\vk \mu 1k{k+1}{h+1}}$ is a
transverse vector. Then

\begin{equation}
E^{\vk \mu 1k{k+1}{h+1}} = Z^{\vk \mu
1k{k+1}{h+1}} = 0
\end{equation}
We conclude then that $\vec{ E} \in {\cal F}$.

\section*{\bf Appendix C}

In this appendix we shall demonstrate: 1) the $\times$-product is closed
with respect to the algebraic constraint and 2) the analytic extension
${\bf X}^\lambda$ of any element of the SeL group satisfies the
constraints. In particular, taking $\lambda = - 1$ we conclude that the
inverse element also satisfies the constraints.

We start considering the algebraic constraint over the element $X_1 \times
X_2$. From (\ref{producto}) we have

\begin{equation}
(X_1 \times X_2)^{\vk \mu 1k{k+1}n} =  \left[ \vec{X_1} \,
+  \vec{X_2}+ \vec{X_1} \times \vec{X_2} \right]^{\vk \mu 1k{k+1}n}
\label{vaproducto}
\end{equation}
The last term of (\ref{vaproducto}) can be written in the form

\begin{equation}
(\vec{X_1} \times \vec{X_2})^{\vk \mu 1k{k+1}n} = \,
\; \sum^{n-1}_{i=1} \, (X_1, X_2)^{\vk \mu 1k{k+1}n}_{i}
\end{equation}
where $(X_1, X_2)^{\vc \mu 1n}_{i} \, \stackrel{\rm def}{=} \, X_1^{\vc \mu
1i}\,X_2^{\vc \mu {i+1}n}$. One can see that

\begin{equation}
(X_1, X_2)^{\vk \mu 1k{k+1}n}_{i} = \sum^{Min(i,k)}_{h=Max(0,k+i-n)}
X_1^{\vk \mu 1h{k+1}{k+i-h}} X_2^{\vk \mu {h+1}k{k+i-h+1}n}
\end{equation}
where we use the following notation, $X^{\vc \mu {i+1}i\vc \mu jm} = X^{\vc
\mu jm\vc \mu {i+1}i} = X^{\vc \mu jm}$. Summing now on i, rearranging the
sums and operating, we obtain

\begin{eqnarray}
(\vec{X_1} \times \vec{X_2})^{\vk \mu 1k{k+1}n} &=& X_1^{\vc \mu 1k}
\left[ \vec{X_1} \times \vec{X_2} +
\vec{X_2} \right]^{\vc \mu {k+1}n} \nonumber \\
&+& X_2^{\vc \mu 1k} \left[ \vec{X_1} \times \vec{X_2}
+ \vec{X_1} \right]^{\vc \mu {k+1}n} \nonumber \\
&+& (\vec{X_1} \times \vec{X_2})^{\vc \mu 1k}
\left[ \vec{X_1} \times \vec{X_2} + \vec{X_1} +
\vec{X_2} \right]^{\vc \mu {k+1}n}
\label{vacontraccion}
\end{eqnarray}

Introducing (\ref{vacontraccion}) in (\ref{vaproducto}) we obtain after a
few direct manipulations

\begin{equation}
(X_1 \times X_2)^{\vk \mu 1k{k+1}n} =
(X_1 \times X_2)^{\vc \mu 1k}(X_1 \times X_2)^{\vc \mu{k+1}n}
\end{equation}

Consider now the analytical extension of an ${\bf X}$. We study first the
differential constraint. One can immediately recognize that $\vec{X_1}
\times \vec{X_2}$ satisfies the differential constraint. The linearity of
the differential constraint assures then that ${\bf X}^\lambda$ obey the
differential identities.

Let us compute the action of the algebraic constraint over ${\bf
X}^\lambda$. We have

\begin{equation}
(\vec {\bf X}^\lambda)^{\vk \mu 1k{k+1}n} = \sum_{i=1}^\infty \c \lambda i
\left[ \, (\vec X)^i \, \right]^{\vk \mu 1k{k+1}n}
\label{penultima}
\end{equation}
The algebraic constrain acting over a $\times$-product of i elements $\vec
X$ produces the following result

\begin{equation}
({\vec X}^i)^{\vk \mu 1k{k+1}n} = \sum_{m=1}^i \, \sum _{j=1}^i \, \c im \c
m{i-j} ({\vec X}^m)^{\vc \mu 1k} \, ({\vec X}^j)^{\vc \mu {k+1}n}
\label{ultima}
\end{equation}
This relation can be obtained directly by induction. Substituting
(\ref{ultima}) in (\ref{penultima}), rearranging the sums and operating we
obtain

\begin{equation}
(\vec {\bf X}^\lambda)^{\vk \mu 1k{k+1}n} =
(\vec {\bf X}^\lambda)^{\vc \mu 1k} \, (\vec {\bf X}^\lambda)^{\vc \mu
{k+1}n}
\end{equation}



\newpage
\begin{thebibliography}{99}

\bibitem{R1} S.Mandelstam, Ann. Phys. 19 (1962) 1.
\bibitem{R2} J.Kogut and L.Susskind, Phys. Rev. D11 (1975) 395.
\bibitem{R3} G.T'Hooft, Nucl. Phys. B153 (1979) 141.
\bibitem{R4} A.M.Polyakov, Phys. Lett. B82 (1979) 247;
Nucl. Phys. B164 (1980) 171.
\bibitem{R5} Yu.M.Makeenko and A.A.Migdal, Nucl. Phys. B188 (1981) 269.
\bibitem{R6} B.Durhuus and P.Olesen, Nucl. Phys. B189 (1981) 406.
\bibitem{R7} A.Ashtekar, New perspectives in canonical gravity (with invited
      contributions) Lecture Notes, Bibliopolis, Naples 1988.
\bibitem{R8} R.Gambini and A.Trias, Phys. Rev. D22 (1980) 1380;
      Nucl. Phys. B278 (1986) 436.
\bibitem{R9} R.Gambini and A.Trias, Phys. Rev. D23 (1981) 553.
\bibitem{R10} R.Gambini and A.Trias, Phys. Rev. D27 (1983) 2935.
\bibitem{R11} R.Gambini and A.Trias, Phys. Rev. D31 (1985) 3144.
\bibitem{R12} R.Gambini and J.Griego, Phys. Lett. B256 (1991) 437.
\bibitem{R13} C.Rovelli and L.Smolin, Nucl. Phys. B331 (1990) 80.
\bibitem{R14} C.Rovelli, Class. and Quantum Grav., 8 (1991) 1613.
\bibitem{R15} A.Ashtekar, Phys. Rev. Lett. 57 (1986) 2244.
\bibitem{R16} L. Smolin, Nonperturbative quantum gravity: the emergence of
discrete structure at the Planck scale, preprint Syracusa/1991.
\bibitem{R17} T.Jacobson and L.Smolin, Nucl. Phys. B299 (1988) 295.
\bibitem{R18} R.Gambini, Phys. Lett. B255 (1991) 180.
\bibitem{R19} B.Brugmann, R.Gambini and J.Pullin, Phys.Rev.Lett.
68 (1992) 431.
\bibitem{R20} E. Witten, Comm. Math. Phys. 121 (1989) 351.
\bibitem{R21} V.F.R. Jones, Bull. Amer. Math. Soc 12 (1985) 103.
\bibitem{R22} P.Freyd et al., Bull. Amer. Math. Soc 12 (1985) 239.
\bibitem{R23} M.Wadati, T.Deguchi and Y.Akutsu, Phys. Rep. 180 (1989) 247.
\bibitem{R24} R.Gambini and L.Leal, Preprint IFFI-91.01 (1991), Montevideo.
\bibitem{SU2} The selfdual property of the connection in the Ashtekar
formalism implies that the SL(2,c) group of Quantum Gravity can be
reduced to an SU(2) group. See for example C.Rovelli, op. cit.
\bibitem{R25} R.Jackiw, Phys. Rev. Lett.41 (1978) 1635.
\bibitem{R26} Y.Ya.Aref'eva, Lett. Math. Phys. 3 (1979) 241.
\bibitem{R27} L.Dolan, Phys. Rev. D22 (1980) 3104.
\bibitem{R28} B.Durhuus and J.M.Leinaas, Phys. Scr. 25 (1982) 504.
\bibitem{R29} X.Fustero, R.Gambini and A.Trias, Phys. Rev. D31 (1985) 3144.
\bibitem{R30} Yu M.Makeenko and A.A.Migdal, Phys. Lett. B.88 (1979) 135.
\bibitem{R31} Ashtekar's private communication.
\bibitem{Guada} E.Guadagnini, M.Martellini and M.Mintchev, Nucl. Phys.B330
(1990) 575.
\bibitem{R33} A.Ashtekar and C.Rovelli, Quantum Faraday lines: loop
representation of the Maxwell theory, Syracuse preprint 1991.
\bibitem{aclaracion} Note that the zero rank component of ${\bf E}$
plays a role analogous to the determinant. For this reason we introduce the
name Special when selecting $E = 1$.
\bibitem{aclare} Notice that ${\it  F}  =  (  0,  \vec{F})$   with
$\vec{F}  \in  {\cal  F}$,  being  ${\cal  F}$  the  vector  space
considered at the  end  of  section  3.1.  The  vanishing  of  the
component  of  rank  zero  makes  this  vector  space   isomorphic
to the one defined by the ${\it F}$'s. For this reason  we  denote
both spaces with the same symbol.
\bibitem{Nori} C.Di Bartolo, F.Nori, R.Gambini and A.Trias,
Lett. Nuovo Cimento 38, (1983) 497.
\bibitem{ARS91} A.Ashtekar, C.Rovelli and L.Smolin, Phys.Rev. D44 (1991) 1740.
\end{thebibliography}
\end{document}